\documentclass[]{interact}

\usepackage{epstopdf}
\usepackage{subfigure}

\usepackage[numbers,sort&compress]{natbib}
\bibpunct[, ]{[}{]}{,}{n}{,}{,}
\usepackage{color}
\theoremstyle{plain}

\theoremstyle{definition}

\theoremstyle{remark}

\begin{document}


\title{Simultaneous detection of quantum oscillations from bulk and topological surface states in metallic Bi$_2$Se$_{2.1}$Te$_{0.9}$}

\author{
\name{Keshav Shrestha\textsuperscript{a,}*\footnote{$^{*}$Email address: keshav.shrestha@inl.gov}, David E. Graf\textsuperscript{b}, Vera Marinova\textsuperscript{c}, Bernd Lorenz\textsuperscript{d}, and Paul C. W. Chu\textsuperscript{d,e}}
\affil{\textsuperscript{a}Idaho National Laboratory, 2525 Fremont Ave., Idaho Falls, ID 83402, USA;  \textsuperscript{b}National High Magnetic Field Laboratory, Florida State University, Tallahassee, FL 32310, USA; \textsuperscript{c}Institute of Optical Materials and Technology, Bulgarian Academy of Sciences, Acad. G. Bontchev Str. 109, Sofia 1113, Bulgaria; \textsuperscript{d}TCSUH and Department of Physics, University of Houston, 3201 Cullen Blvd., Houston, Texas 77204, USA; \textsuperscript{e}Lawrence Berkeley National Laboratory, 1 Cyclotron Road, Berkeley, California 94720, USA}
}

\maketitle

\begin{abstract}
Shubnikov-de Haas (SdH) oscillations in metallic Bi$_2$Se$_{2.1}$Te$_{0.9}$ are studied in magnetic fields up to 35 Tesla. It is demonstrated that two characteristic frequencies determine the quantum oscillations of the conductivity. Angle dependent measurements and calculations of the Berry phase show that the two frequencies $F_1$ and $F_2$ describe oscillations from surface and bulk carriers, respectively. At low magnetic fields, only SdH oscillation from topological surface states can be detected whereas at high magnetic field the bulk oscillations dominate. The origin of the separation of bulk and surface SdH oscillations into different magnetic field ranges is revealed in the difference of the cyclotron masses $m_c$. The bulk $m_c$ is nearly three times larger than the surface cyclotron mass resulting in a stronger attenuation of the bulk oscillation amplitude upon decreasing magnetic field. This makes it possible to detect and characterize the surface SdH oscillations in the low-field range and the bulk oscillations at high magnetic fields.
\end{abstract}

\begin{keywords}
Topological insulator; Shubnilov-de Haas oscillations; surface states; bulk states
\end{keywords}

\section{Introduction}
Topological surface states in systems with strong spin-orbit interactions have become the focus of interest in recent years \cite{hasan:10,qi:11,ando:13}. The nontrivial topology of the respective electronic states results in a number of novel quantum phenomena, for example, the surface of a topological insulator (TI) has to be conducting since the transition from a nontrivial insulator to the vacuum or an ordinary insulator demands the electronic surface states to be gapless. The surface states are usually protected by symmetry, e.g. the time reversal symmetry in a three-dimensional TI. The intrinsic characteristics of topological surface states is the Dirac-like dispersion and their two-dimensional character, both properties can be experimentally verified.

While the band structure and the existence of electronic states with a Dirac-like dispersion filling the gap between the valence and conduction bands can be verified in angle resolved photoelectron spectroscopy (ARPES), the electrical transport properties of the surface states and the associated quantum oscillations in magnetic fields (Shubnikov-de Haas effect) are frequently utilized to prove the topological surface states in various systems. This method works very well as long as the bulk states of the TI are insulating, i.e. the Fermi energy falls into the bulk band gap \cite{taskin:11,eguchi:14}. However, many topological systems are conducting in the bulk due to defects resulting in a shift of the Fermi energy into the conduction or valence bands. A typical example is the Bi$_2$Se$_3$ - Bi$_2$Te$_3$ system \cite{xia:10,chen:10,hsieh:10} where it was also found that bulk conduction interfering with surface conduction made it difficult, if not impossible, to detect the specific signature of the surface states in Shubnikov-de Haas (SdH) oscillations of the conductivity \cite{qu:11,analytis:11,eto:11,cao:13}.

However, in a recent study of metallic Bi$_2$Se$_{2.1}$Te$_{0.9}$ we found the distinct signature of topological surface states in SdH oscillations by investigating the dependence of the quantum oscillations on the angle of the magnetic field with the surface of the crystal as well as the Berry phase which was found consistent with the Dirac nature of the conducting particles (holes) \cite{shrestha:14}. Surprisingly, in the magnetic field range up to 7 Tesla, no SdH oscillations from bulk carriers have been observed in this study although the bulk conduction turned out to be metallic to the lowest temperatures and the total carrier density was relatively high, leading to the speculation that the quantum oscillations from bulk carriers had been strongly suppressed at low fields but might be found at higher magnetic fields.

In this work we extend the magnetic field range to 35 Tesla and study the SdH oscillation spectrum in hole-like metallic Bi$_2$Se$_{2.1}$Te$_{0.9}$. We prove that indeed SdH oscillations from bulk carriers dominate in the high-field range and are attenuated at lower fields. This enables us to separate bulk and surface states, determine the relevant parameters, and explain the interference and possible separation of surface and bulk quantum oscillations.

\section{Experimental}
The growth of the single crystals of Bi$_2$Se$_{2.1}$Te$_{0.9}$ was achieved by utilizing a modified Bridgman technique with high purity starting materials Bi (99.9999\%), Se (99.9999\%), and Te (99.9999\%). The mixture, enclosed in quartz ampoules, was molten at 875 $^\circ$C and kept at this temperature for 2 days. The melt was slowly cooled to 670 $^\circ$C at a rate of 0.5 $^\circ$C/h. The crystals were finally cooled to room temperature at a speed of 10 $^\circ$C/h. Platelet-like crystals of typical size 5 mm x 3 mm x 0.1 mm have been extracted from the synthesis product. All crystals prepared for transport measurements have been cleaved to provide fresh and clean surfaces.

Magnetotransport measurements have been conducted using a lock-in technique at the National High Magnetic Field Laboratory (NHMFL) in Tallahassee, FL. Six gold contacts were sputtered onto one surface of the crystal to conduct longitudinal (resistance) and transverse (Hall) measurements. Platinum wires were attached to the gold pads with silver paint. The sample was mounted on a rotating platform which allowed for positioning the sample at different angles with the magnetic field. The platform, mounted in a $^3$He cryostat (Oxford), was inserted into the 32 mm bore of a resistive magnet with a maximum field of 35 Tesla.

\section{Results and Discussion}
\subsection{Shubnikov-de Haas oscillations}
The metallic character of the bulk conductivity is demonstrated in Fig. 1. The continuous decrease of the resistivity $\varrho_{xx}(T)$ upon decreasing temperature and the positive slope of the Hall resistance $R_{xy}(B)$ (shown in the inset to Fig. 1) prove that the charge carriers are holes and the Fermi energy is positioned below the top of the valence band, consistent with other crystals from the same growth batch \cite{shrestha:14}. From the Hall data measured at 5 K (Fig. 1), the bulk hole carrier density is determined as $n_{bulk}=$1.3$\times$10$^{18}$ cm$^{-3}$. This number is in good agreement with the carrier density of a similar crystal of the same chemical composition \cite{shrestha:14}.

 Longitudinal and Hall resistance of Bi$_2$Se$_{2.1}$Te$_{0.9}$ are measured in high magnetic field up to 35 Tesla at the NHMFL, as shown in figure 2. Both $R_{xx}(B)$ and $R_{xy}(B)$ show quantum oscillations at high magnetic fields. The quantum oscillations in longitudinal and Hall components have a same frequency but have a phase difference of 90$^{o}$. Since in our data it is easier to subtract a background signal in Hall component than that in longitudinal, we have taken Fourier transform of the Hall signal for frequency analyses. The oscillatory part $\Delta R_{xy}$, obtained after subtracting a smooth polynomial background, is shown in figure 3. It is obvious from Fig. 3 that the data cannot be described by an oscillation with one single frequency only, but rather by a superposition of different frequencies. This is confirmed by analyzing the Fourier transform (FFT) of the data from Fig. 3. The FFT's of Fig. 3 in different field ranges are displayed in Fig. 4.

\begin{figure}
\begin{center}
\includegraphics[angle=0,width=4in]{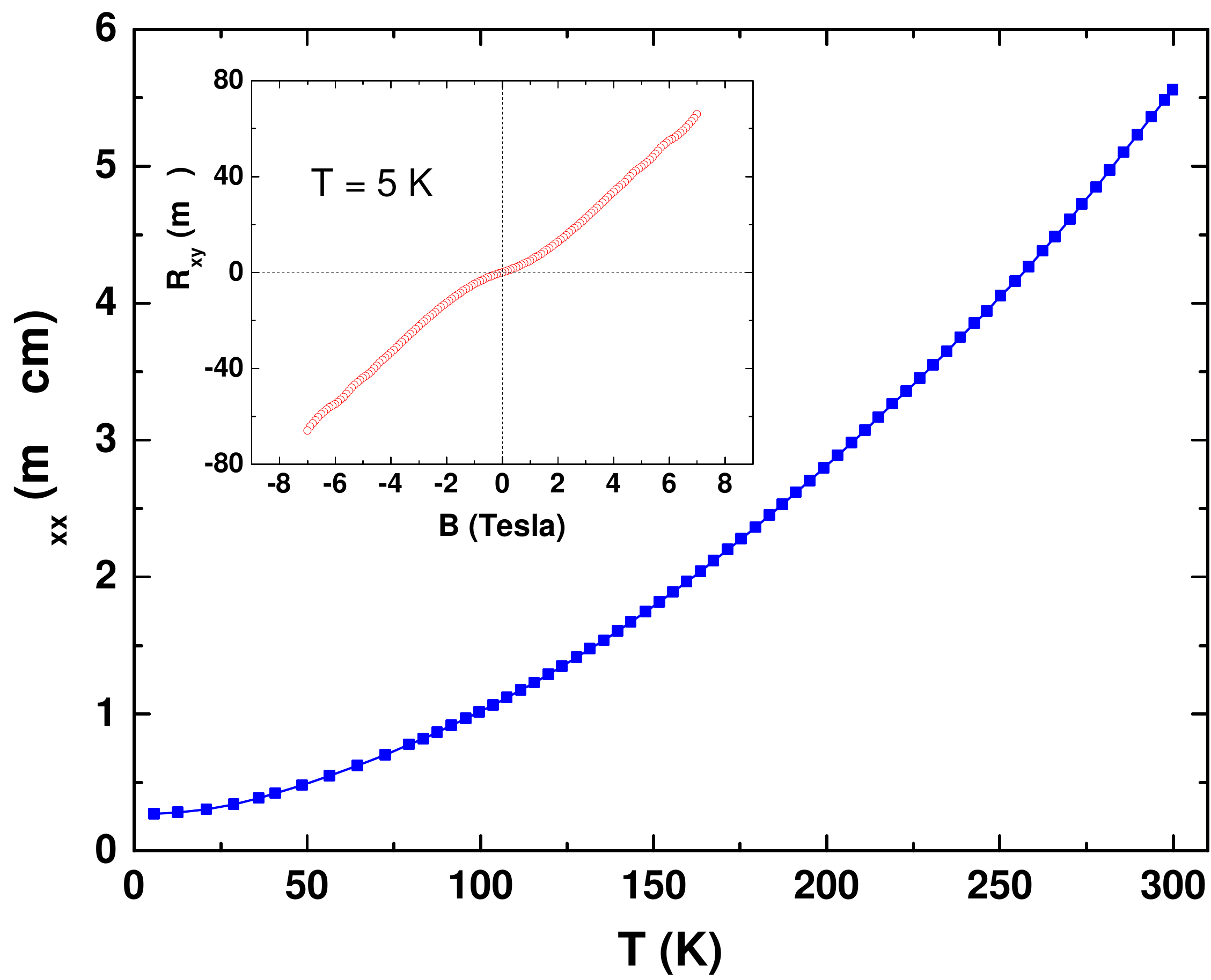}
\caption{Longitudinal resistivity $\varrho_{xx}(T)$ of Bi$_2$Se$_{2.1}$Te$_{0.9}$ versus temperature. The inset shows the transverse resistance $R_{xy}$ at 5 K as function of the magnetic field.}
\end{center}
\end{figure}

\begin{figure}
\begin{center}
\includegraphics[angle=0,width=5in]{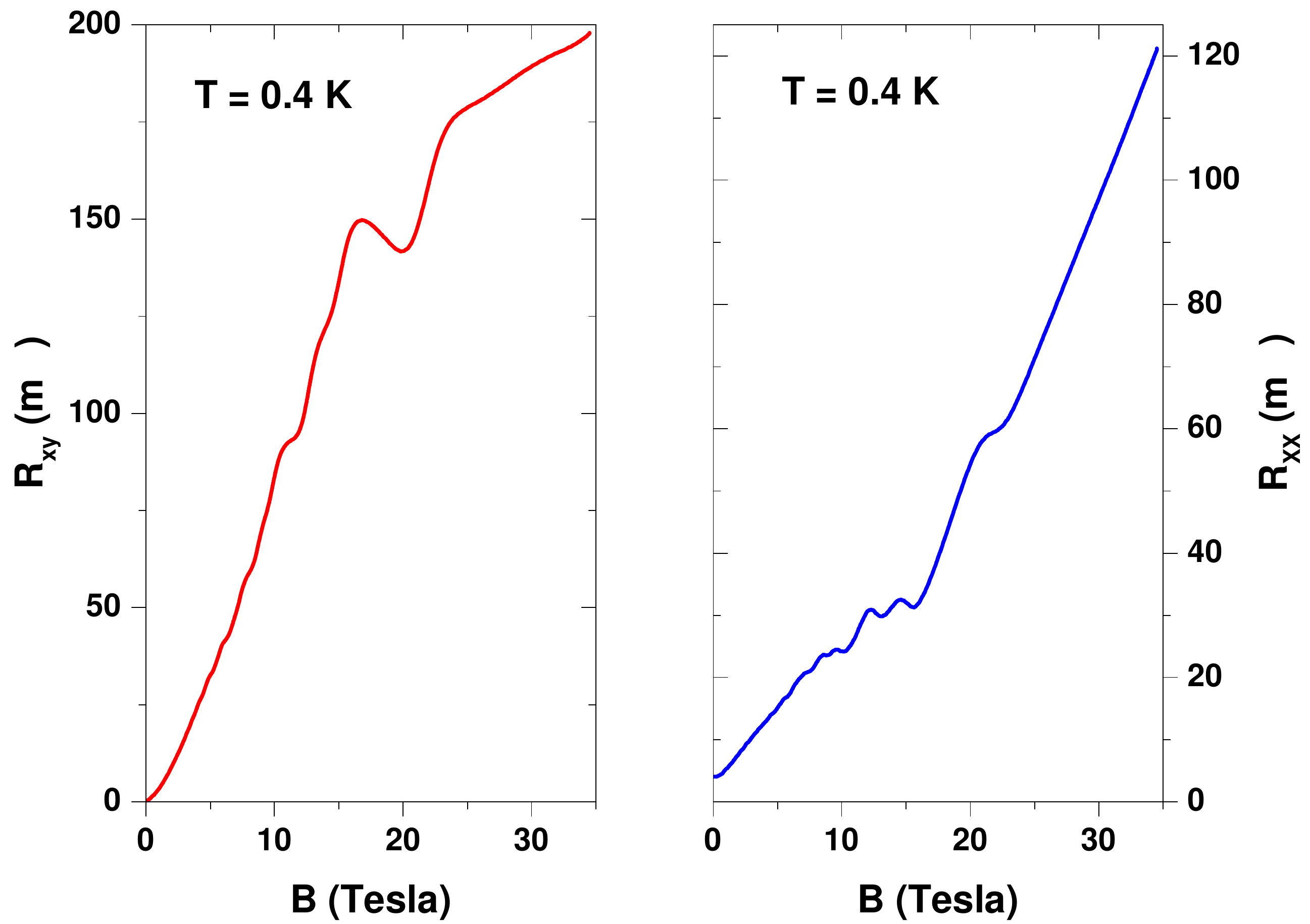}
\caption{Longitudinal and transverse magentoresistance of Bi$_2$Se$_{2.1}$Te$_{0.9}$ in the magnetic fields up to 35 Tesla, measured at $T$ = 0.4 K. Both longitudinal and Hall resistances show clear quantum oscillations at higher fields.}
\end{center}
\end{figure}

\begin{figure}
\begin{center}
\includegraphics[angle=0,width=4in]{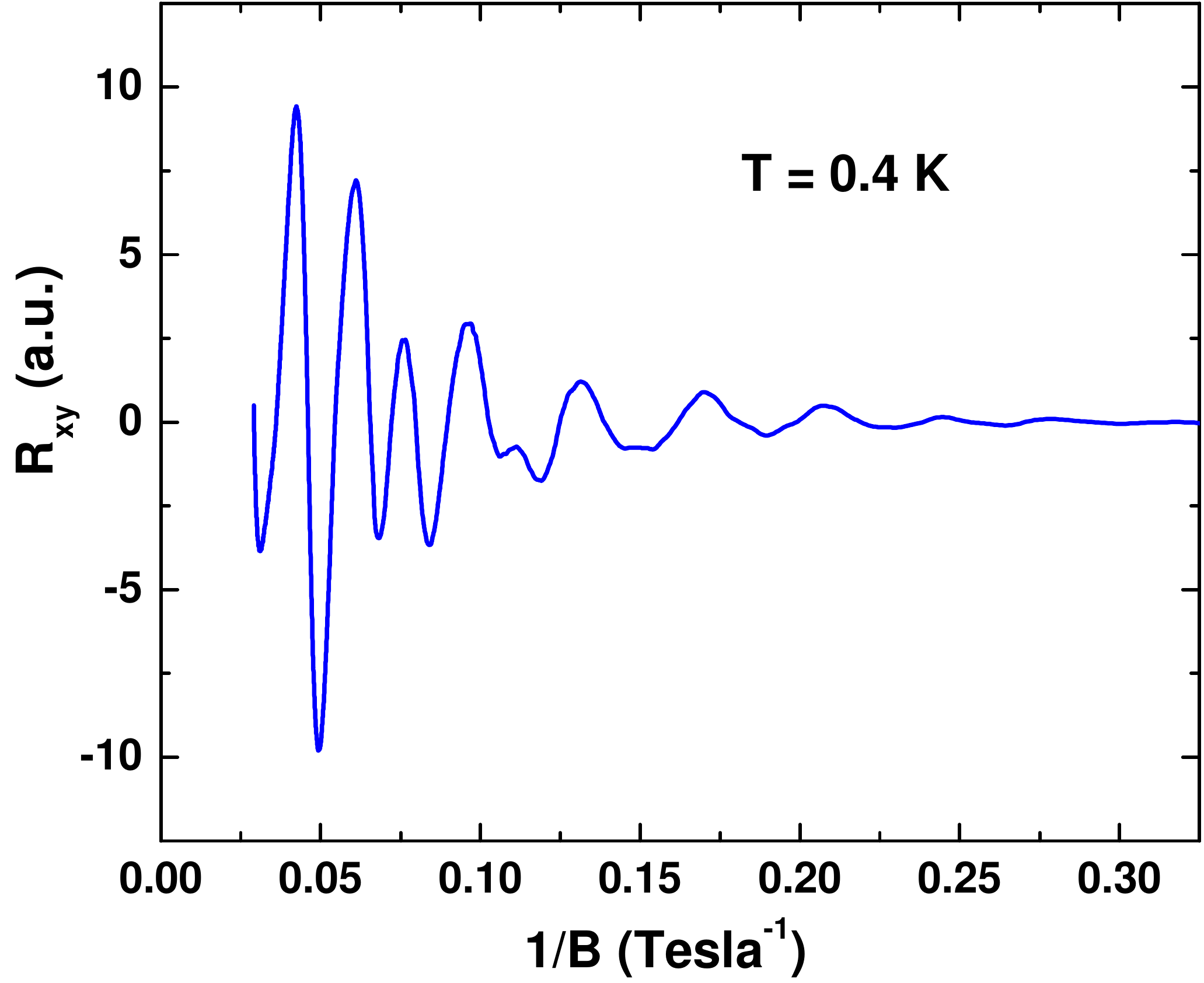}
\caption{Shubnikov-de Haas oscillation of the transverse magnetoresistance of Bi$_2$Se$_{2.1}$Te$_{0.9}$. The oscillatory part $\Delta R_{xy}$ is plotted versus the inverse field.}
\end{center}
\end{figure}
\begin{figure}
\begin{center}
\includegraphics[angle=0,width=4.5in]{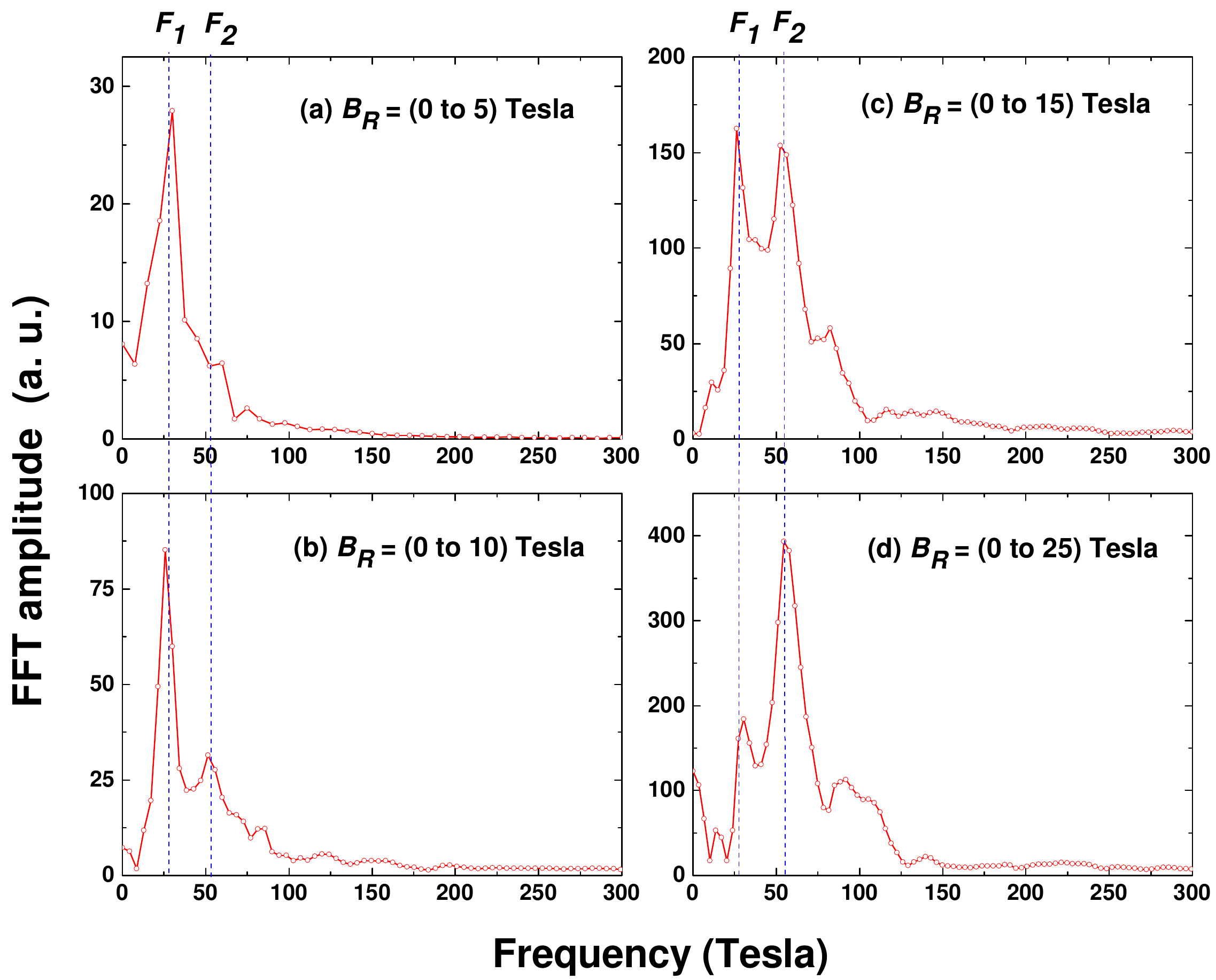}
\caption{Fourier transform of $\Delta R_{xy}$ in different field ranges. The cutoff field is indicated in the graphs. The SdH oscillations with frequencies $F_1$ and $F_2$ are dominant at low and high magnetic fields, respectively.}
\end{center}
\end{figure}

Two frequencies, $F_1\approx$ 26 Tesla and $F_2\approx$ 55 Tesla, dominate the oscillating behavior of $R_{xy}$. Since $F_2$ is nearly twice $F_1$, both frequencies could be the first and second harmonic of the same oscillation, as observed in other compounds \cite{taskin:09,jalan:10,taskin:11,li:14}. However, the relative weight of the oscillations with $F_1$ and $F_2$ depends on the magnetic field range suggesting that $F_2$ is not likely to be the second harmonic of $F_1$. This is supported in our analysis of the SdH oscillations proving its bulk origin from both the angle dependence measurements and Berry phase calculations. The SdH oscillation with the lower frequency $F_1$ dominates in the low-field range whereas the higher frequency $F_2$ is stronger at higher magnetic fields. This is demonstrated in Fig. 4 where the Fourier transform of the data from Fig. 4 is shown for various field ranges, $B_R$ as indicated in the figures. In the low-field $B_R$ = (0 to 5) Tesla, the FFT exhibits only one peak at frequency $F_1$ (Fig. 4a). This is similar to and consistent with the earlier work that was limited to magnetic fields below 7 Tesla \cite{shrestha:14}. With increasing magnetic field, a second peak at $F_2$ develops (Fig. 4b) and for fields up to 15 tesla both peaks have about the same magnitude (Fig. 4c). With further increasing field, the $F_2$ peak becomes dominant (Fig. 4d).

The development of the two peaks shown in Fig. 4 prove that $F_1$ and $F_2$ characterize SdH oscillations of different origin. In our previous communication, we have shown that the low-frequency oscillation ($F_1$) arises from topological surface states, but the origin of the second frequency observed at higher fields is not clear. It appears conceivable to attribute the $F_2$ frequency to bulk SdH oscillations, as conjectured earlier \cite{shrestha:14}. To study the properties of the $F_2$ oscillation it has to be resolved separately, without the interference from the surface state oscillations ($F_1$). This can be achieved by analyzing the high-field data above 10 Tesla. Fig. 5 shows that the FFT of the data above 10 Tesla exhibits only one pronounced peak at frequency $F_2$, i.e. the contribution from surface oscillations is largely eliminated.

\begin{figure}
\begin{center}
\includegraphics[angle=0,width=3in]{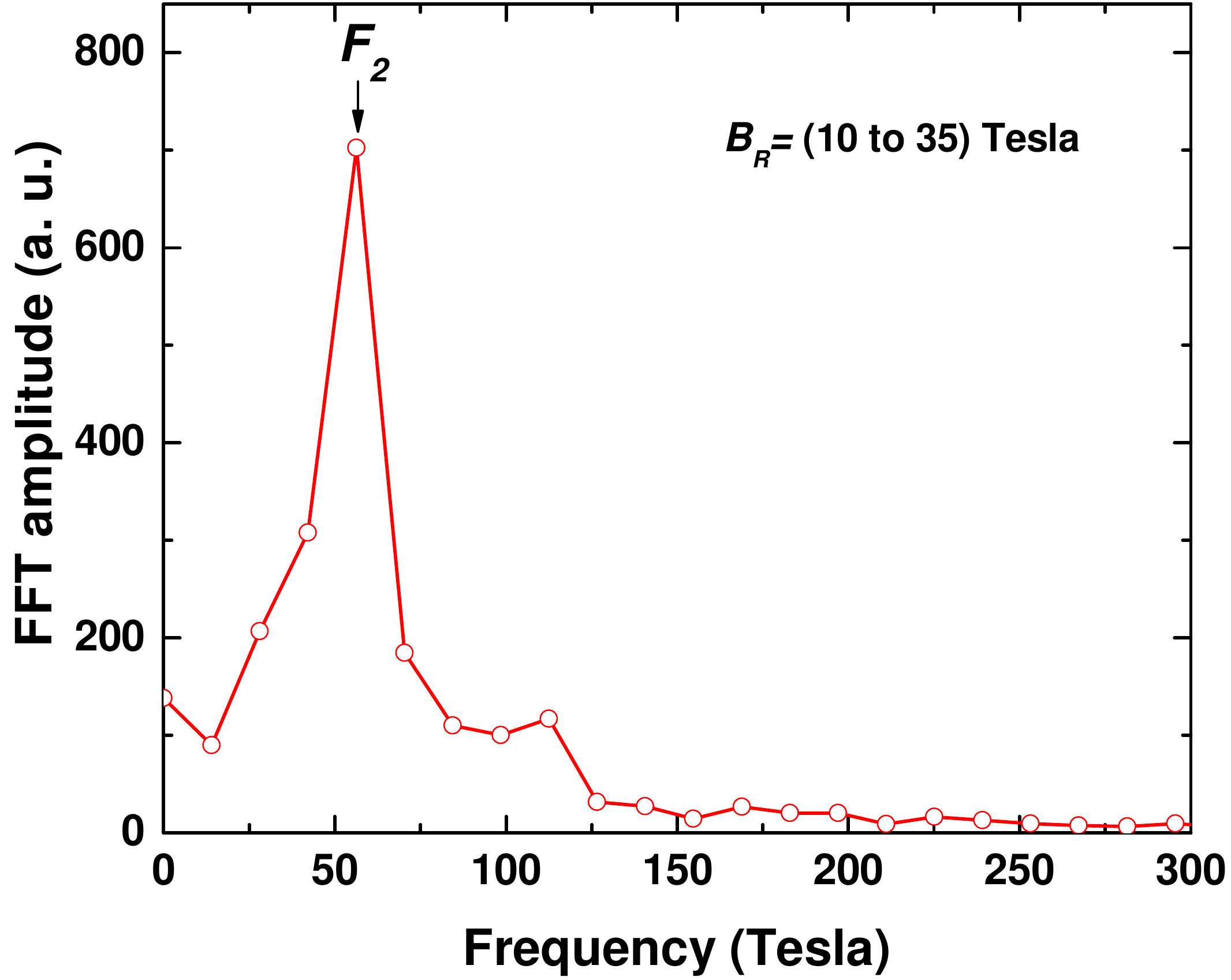}
\caption{Fourier transform of $\Delta R_{xy}$ in the high-field range, between 10 and 35 Tesla. Only one peak is observed at frequency $F_2$.}
\end{center}
\end{figure}

\subsection{Angle dependence of SdH oscillations}
SdH oscillations from bulk and surface states can be distinguished by measuring the dependence on the angle with the magnetic field. Surface oscillations of $\Delta R_{xy}$ or $\Delta R_{xx}$ are expected to be periodic if plotted as function of the inverse normal component, 1/$B_\perp$, of the field with respect to the surface. If the field angle $\Theta$ to the normal of the surface changes, the position of the oscillation frequency follows a 1/cos$\Theta$ scaling, due to the strictly two-dimensional character of the surface conduction \cite{ren:10,li:14}. For bulk conduction, however, the SdH frequency will not follow the 1/cos$\Theta$ scaling, but it may still show a minor angle dependence if the Fermi surface geometry is anisotropic.

The angle dependent measurements have been conducted over the whole field range up to 35 Tesla and angles between 0$^\circ$ and 70$^\circ$. The Fourier transform to determine $F_1$ and $F_2$ at different field angles was calculated using the data measured at various angles $\Theta$, similar to the data for $\Theta$=0$^\circ$ in Fig. 3. As shown in Fig. 6, the frequency $F_1$ scales well with 1/cos($\Theta$) (dashed line in Fig. 6a) indicating that this conduction channel is two-dimensional. The $F_2$ oscillation, however, changes only very little with the angle $\Theta$ and is therefore attributed to the bulk conduction channel (Fig. 6b).

\begin{figure}
\begin{center}
\includegraphics[angle=0,width=4in]{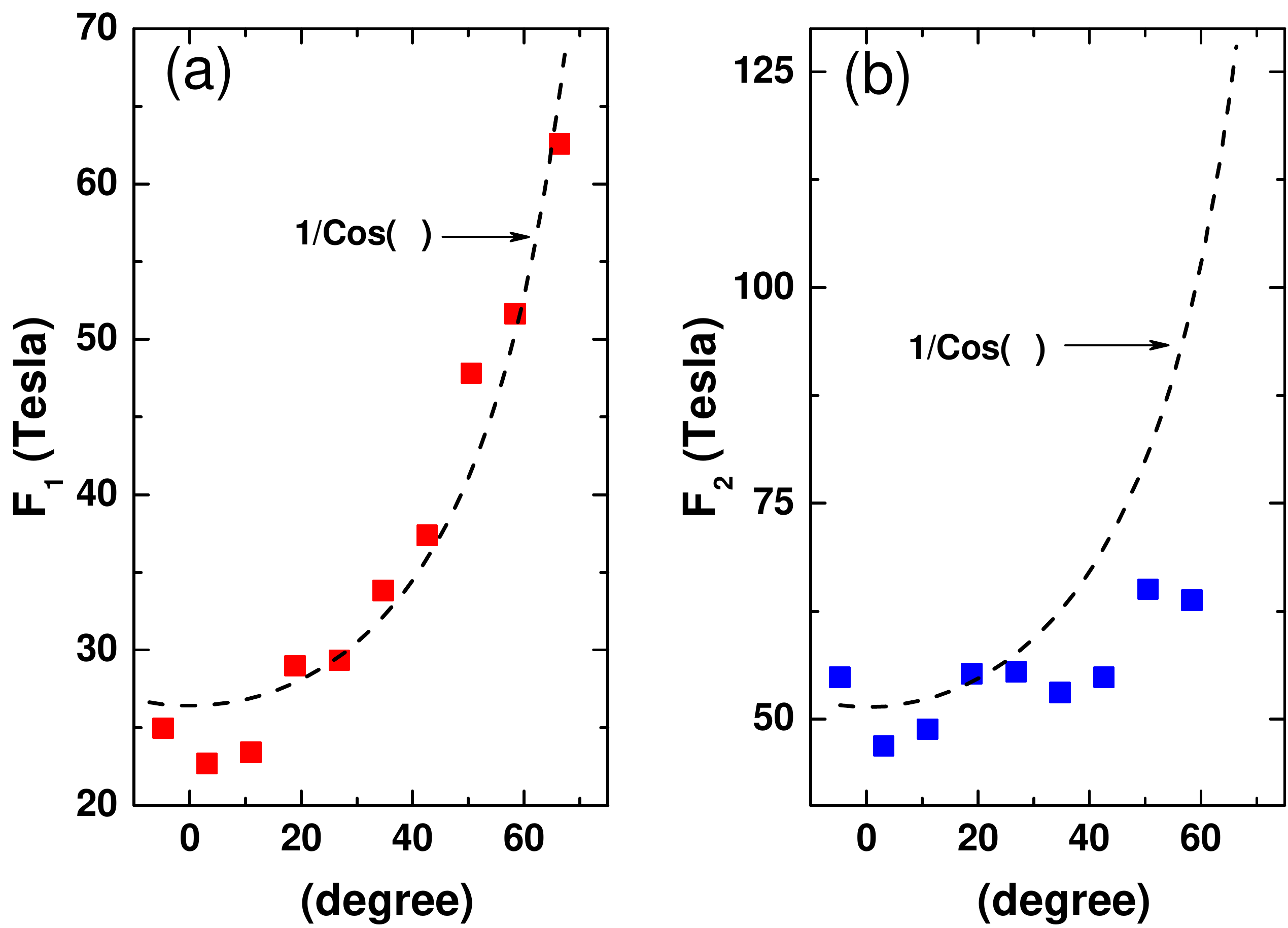}
\caption{Angle dependence of the SdH oscillation frequencies $F_1$ (a) and $F_2$ (b). Only $F_1$ follows the 1/cos($\Theta$) scaling for surface conduction (dashed line).}
\end{center}
\end{figure}

It should be noted that there is a small shoulder in the Fourier transform near 100 Tesla visible in Figs. 4d and 5. This shoulder develops into a peak with frequency $F_3\approx$ 90 to 100 Tesla with increasing angle $\Theta$. This additional peak is attributed to another section of the bulk Fermi surface which contributes to the SdH oscillations only at higher angles $\Theta$. Since the value of $F_3$ is nearly independent of the angle $\Theta$, it cannot arise from surface states.

\subsection{Berry phase}
The angle dependent transport data discussed so far lead to the conclusion that SdH oscillations from bulk and topological surface states can be measured simultaneously and resolved separately in different magnetic field ranges. The conclusion is further supported by an analysis of the Berry phase which distinguishes the nature of the charge carriers. The charge carriers of the topologically nontrivial surface states with a Dirac dispersion are expected to have a Berry phase $\beta=$1/2, in contrast to the bulk carriers with a Berry phase of zero. $\beta$ can be determined from the Landau level fan diagram \cite{ando:13, shrestha1:10}.

It has been shown that the SdH oscillations of the conductivity $\Delta\sigma$, in contrast to oscillations of the resistivity $\Delta\varrho$, provide a more accurate determination of the Berry phase \cite{ando:13}. To determine the nature of the charge carriers in the high-field range (with oscillation frequency $F_2$), we have to evaluate the SdH oscillations at sufficiently high fields, cutting off the low-field data, to eliminate any interference from the surface oscillations. It will be shown below, that the crossover from surface ($F_1$) to dominantly bulk ($F_2$) oscillations takes place at $B_c\approx$14 Tesla. Fig. 7 shows the oscillating part of longitudinal conductivity $\sigma_{xx}(B)$ above 13 Tesla, calculated using the formula, $\sigma_{xx}=\varrho_{xx}/(\varrho_{xx}^2+\varrho_{xy}^2)$. The vertical dashed lines indicate the positions of maxima and minima. It was shown that the minima and maxima of the $\Delta\sigma_{xx}$ correspond to the integer and half-integer numbers of $n$, respectively \cite{qu:11}. The Landau level fan diagram is shown in the inset of Fig. 7. The plot $n$ vs. 1/$B_n$ reveals a linear relation given by $F/B_n-\beta=n-1$ and the value $n$ obtained from the extrapolation 1/$B_n\rightarrow$0 is very close to 1. Accordingly, the linear fit determines the Berry phase as $\beta$=0.038$\pm$0.071. This value is consistent with the bulk nature of the charge carriers which give rise to the SdH oscillations in the high-field range \cite{ando:13}, in agreement with the weak angle dependence (Fig. 6b).
\begin{figure}
\begin{center}
\includegraphics[angle=0,width=4.5in]{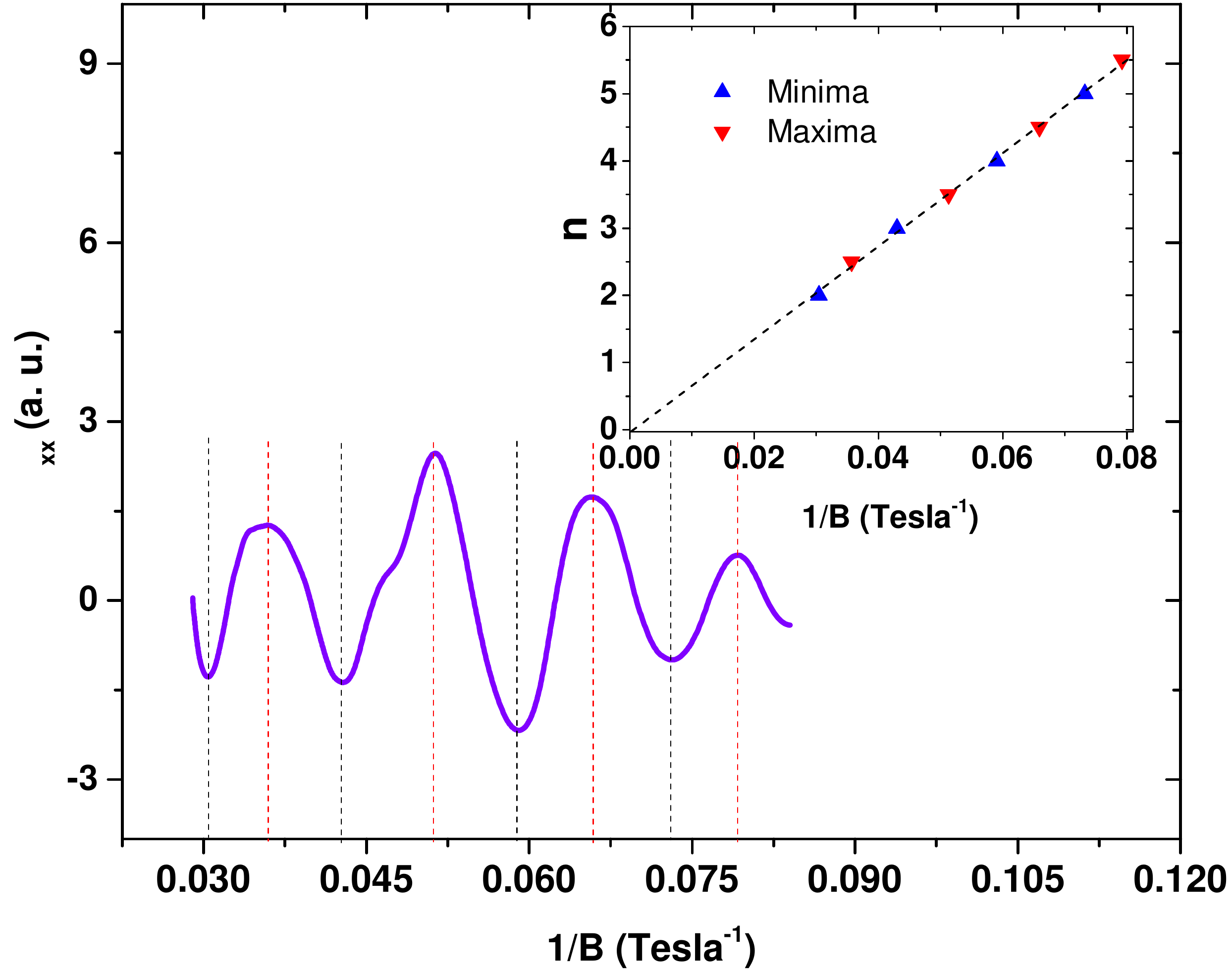}
\caption{SdH oscillation of the conductivity $\Delta\sigma_{xx}$ vs. $1/B$ in the high-field range at $T$ = 5 K. The positions of maxima and minima are indicated by vertical dashed lines. The inset shows the Landau level fan diagram. The linear extrapolation 1/$B_n\rightarrow$ 0 determines the Berry phase $\beta$. Up and down triangles represent the positions of the $\Delta\sigma_{xx}$ minima and maxima, respectively.}
\end{center}
\end{figure}

For comparison, the Berry phase of the surface carriers is determined from the low-field data, $B<$ 7 Tesla. In this field range, the SdH oscillations are pronounced in the second derivative of $\sigma_{xx}$ with respect to the inverse field 1/$B$ (see Fig. 8). Here the maxima have to be assigned to integer values of the Landau level index $n$, as labeled in Fig. 8. The linear extrapolation of the Landau level fan plot (inset to Fig. 8) to 1/$B\rightarrow0$ reveals a value of $n_0$=0.45 corresponding to a Berry phase of $\beta$=0.55$\pm$0.06. This value is in very good agreement with the earlier data for a similar crystal of Bi$_2$Se$_{2.1}$Te$_{0.9}$ \cite{shrestha:14}. The value of $\beta$ close to 0.5 proves the Dirac nature of the topological surface carriers.

\begin{figure}
\begin{center}
\includegraphics[angle=0,width=4.5in]{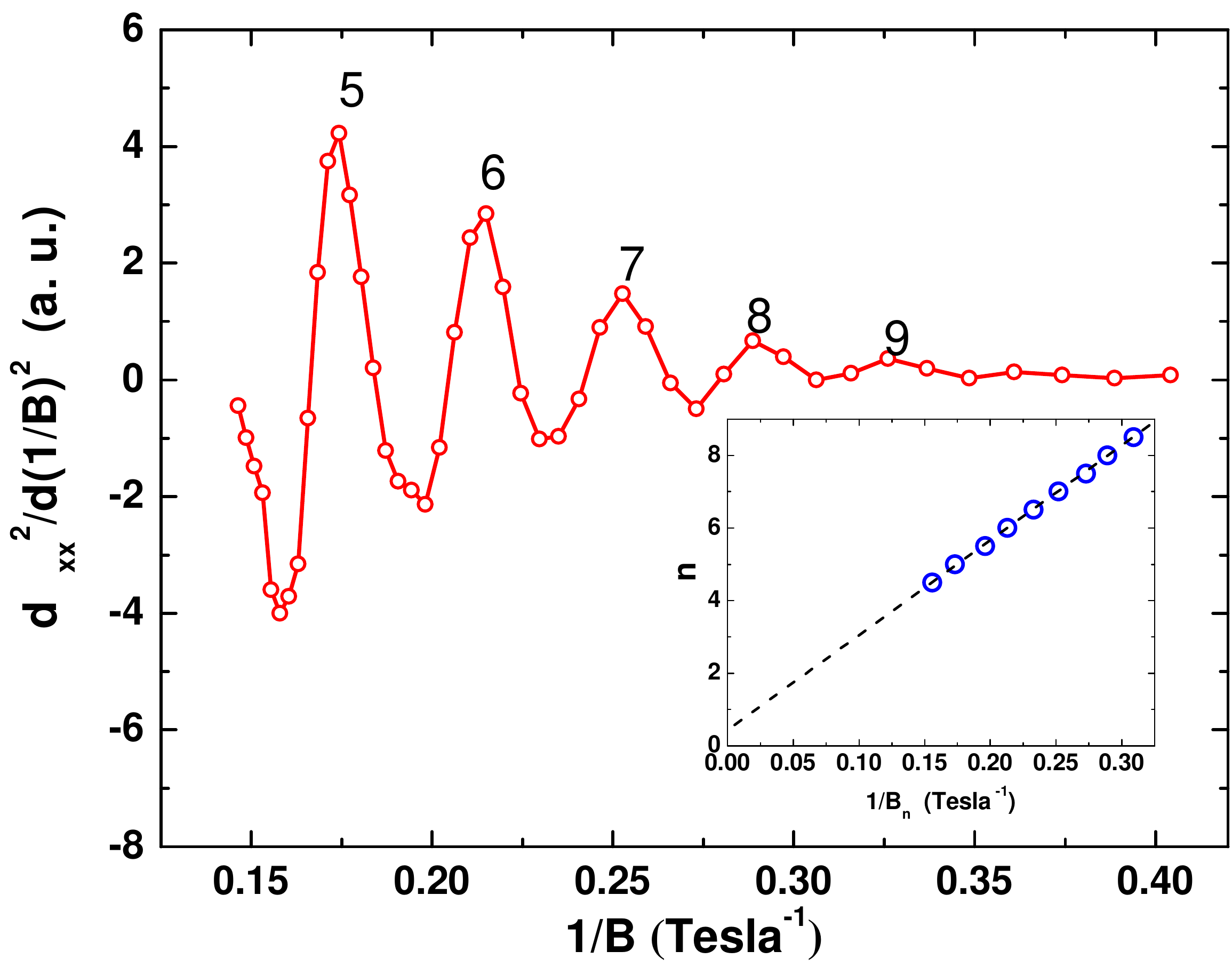}
\caption{SdH oscillation of the conductivity $\sigma_{xx}$ vs. $B^{-1}$ in the low-field range ($B<$ 7 Tesla) at $T$=5 K. Shown is the second derivative where the maxima correspond to integer values of $n$, as labeled. The inset shows the Landau level fan diagram. The linear extrapolation 1/$B_n\rightarrow$ 0 (dashed line) determines the Berry phase $\beta$=0.55.}
\end{center}
\end{figure}

\subsection{Lifshitz-Kosevich analysis}
The separation of SdH oscillations arising from topological surface and trivial bulk states into low- and high-field ranges, respectively, needs to be understood. To this end, the microscopic parameters defining the quantum oscillations have to be determined. This can be achieved through the Lifshitz-Kosevich (LK) analysis of the SdH oscillations of $\Delta R_{xx}$ measured at different temperatures. According to the LK theory, the amplitude of the SdH oscillation of $\Delta R_{xx}$ is expressed as function of temperature and magnetic field: \cite{qu:10,ando:13}

\begin{equation}
\Delta R(T,B) = \Delta R_0 e^{-\lambda_D(B)} {\lambda(T/B)\over {sinh[\lambda(T/B)]}}
\end{equation}

with

\begin{equation}
\lambda_D(B) = {2\pi^2k_B\over {\hbar e}} m_c {T_D\over B}
\end{equation}

\begin{equation}
\lambda(T/B) = {2\pi^2k_B\over {\hbar e}} m_c {T\over B}
\end{equation}

The first term in equ. (1), $\Delta R_0$, is the amplitude of the oscillation in the high-field limit 1/$B\rightarrow$0. The next term is the Dingle factor representing the exponential decrease of $\Delta R$ with decreasing field $B$. The last term describes the attenuation of $\Delta R$ with increasing temperature $T$. $m_c$ is the cyclotron mass of the charge carriers and $T_D$ is the Dingle temperature which is related to the inverse life time of the carriers.

There are only three fit parameters in equs. (1) to (3), $\Delta R_0$, $m_c$, and $T_D$, which can be determined for a specific oscillation by analyzing the field and temperature dependencies of $\Delta R(T,B)$. Fig. 9a shows the SdH oscillations of $\Delta R_{xx}$ in the high-field range at different temperatures. The temperature dependence of $\Delta R$ is solely determined by the $\lambda/sinh\lambda$ term in equ. (1). Fitting this expression to the data at different constant magnetic fields, e.g. at 25 Tesla shown in Fig. 10, allows for the determination of the Landau level spacing $\Delta E_N(B)=\hbar eB/m_c$ and the cyclotron mass $m_c$ from the slope of the plot $\Delta E$ vs. $B$ in the lower inset of Fig. 10. For the high-field (bulk) oscillations we obtain $m_c=0.34 m_e$ ($m_e$ is the bare electron mass).

\begin{figure}
\begin{center}
\includegraphics[angle=0,width=4.5in]{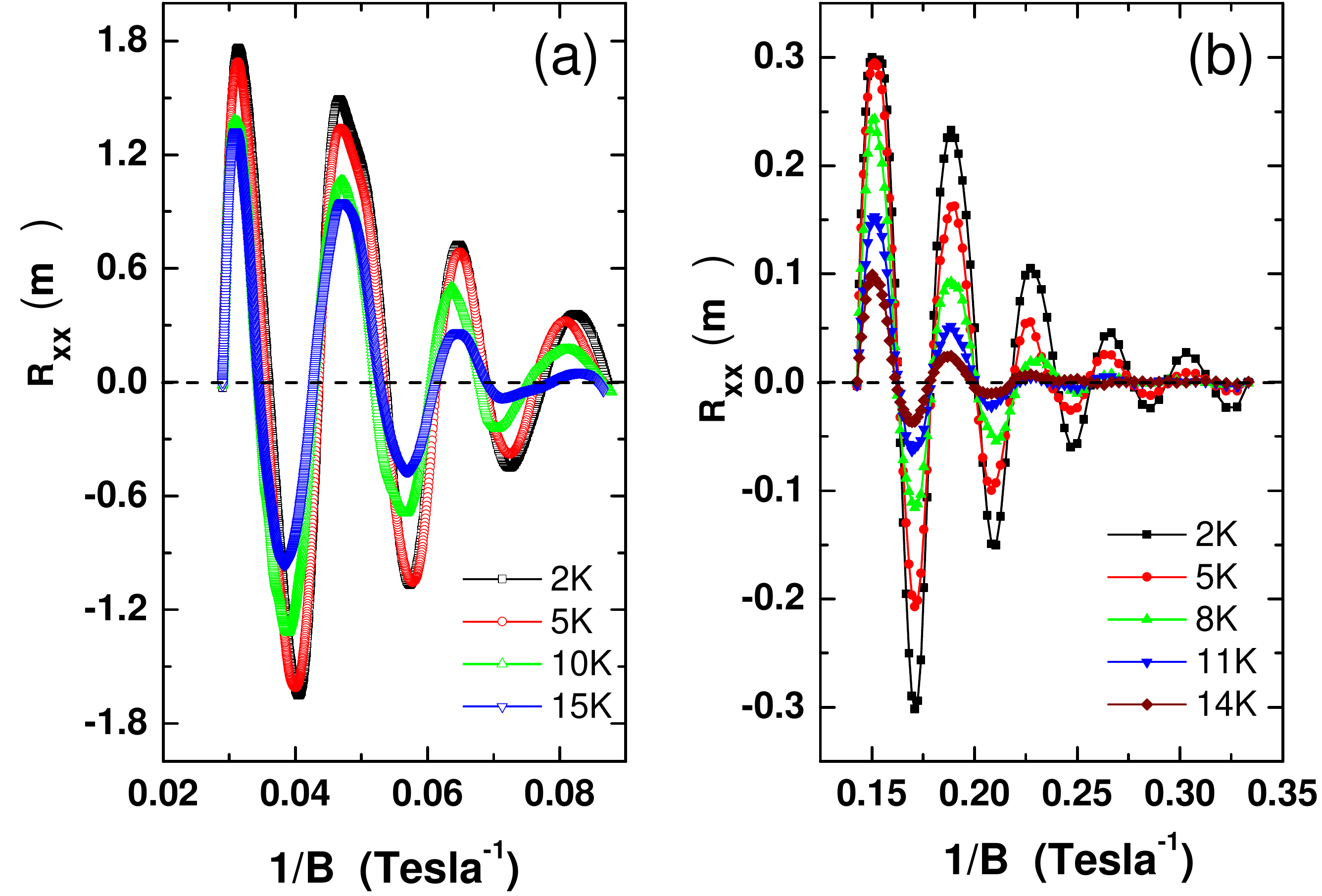}
\caption{SdH oscillation of $\Delta R_{xx}$ vs. $B^{-1}$ in the high-field (a) and low-field (b) ranges at different temperatures between 2 and 15 K.}
\end{center}
\end{figure}

\begin{figure}
\begin{center}
\includegraphics[angle=0,width=4.5in]{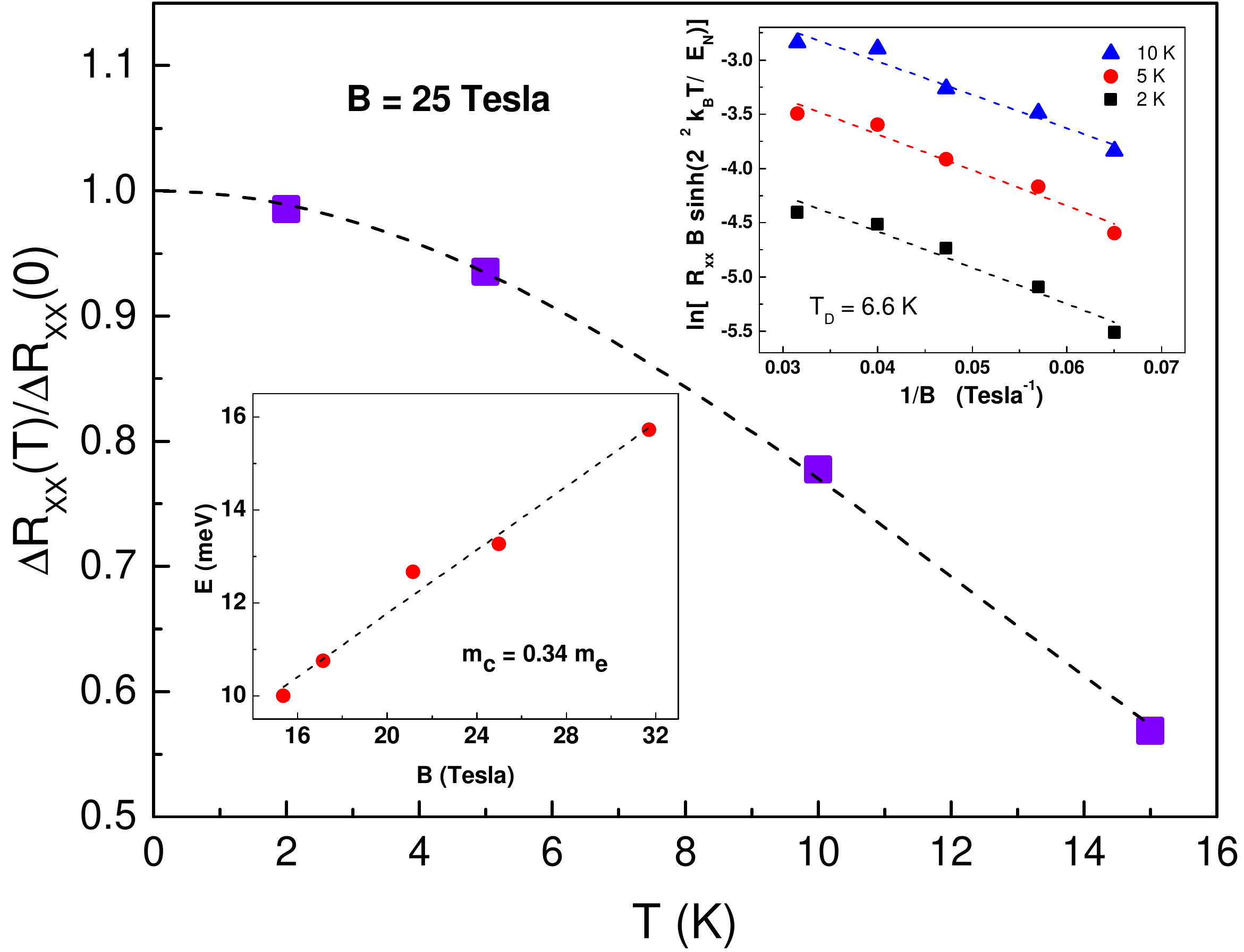}
\caption{LK analysis of the high-field SdH oscillation with frequency $F_2$. Main panel: Temperature dependence of the amplitude $\Delta R_{xx}$ at 25 Tesla. The line is a fit to the LK formula, equ. (1). The lower inset shows the Landau level spacing $\Delta E$ as function of magnetic field $B$. The cyclotron mass $m_c$ = 0.34 $m_e$ is determined from the slope of $\Delta E(B)$. The upper inset is the semi-logarithmic Dingle plot from which the Dingle temperature $T_D$ = 6.6 K is obtained.}
\end{center}
\end{figure}
\begin{figure}
\begin{center}
\includegraphics[angle=0,width=4.5in]{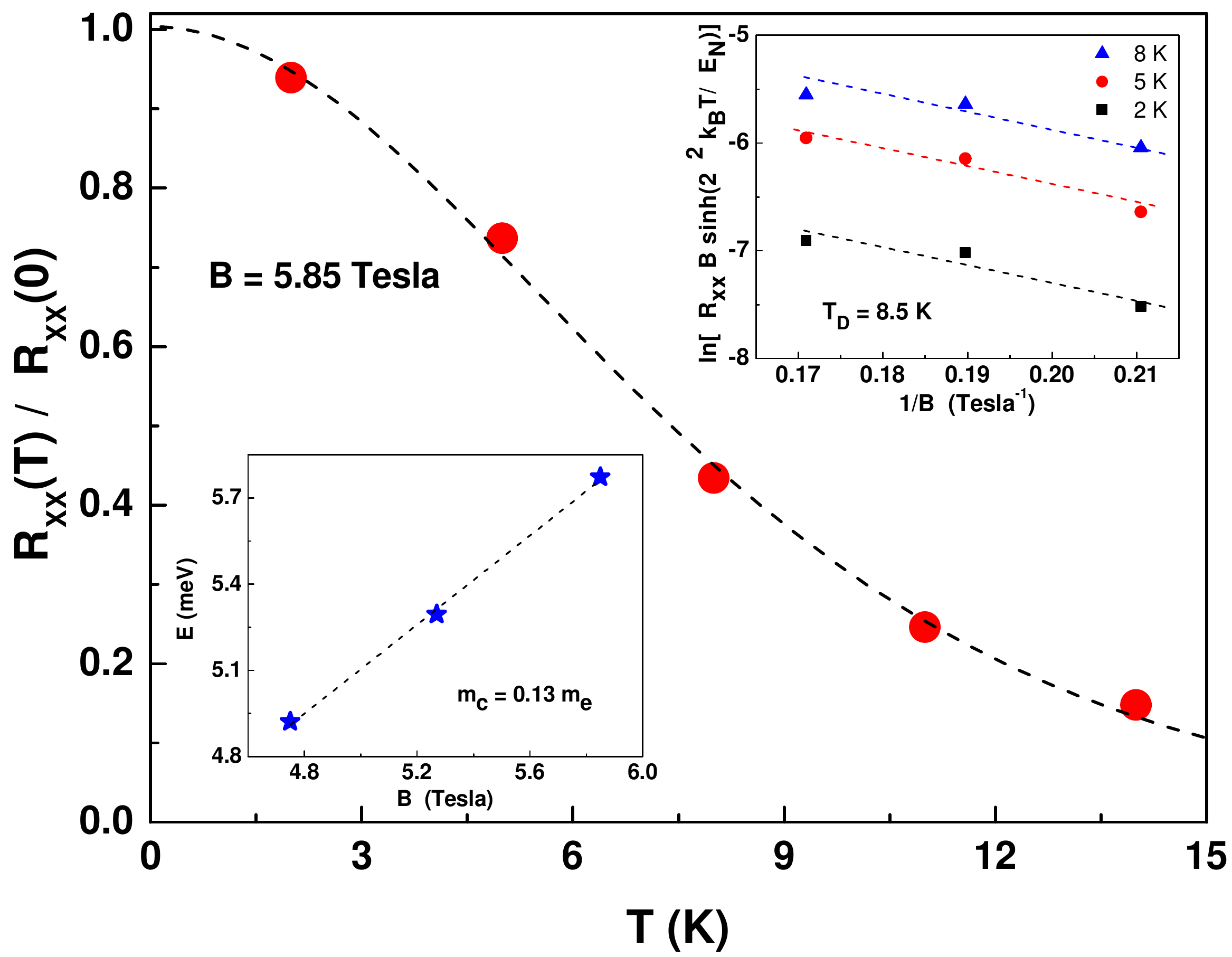}
\caption{LK analysis of the low-field SdH oscillation with frequency $F_1$. Main panel: Temperature dependence of the amplitude $\Delta R_{xx}$ at 5.85 Tesla. The line is a fit to the LK formula equ. (1). The lower inset shows the Landau level spacing $\Delta E$ as function of magnetic field $B$. The cyclotron mass $m_c$ = 0.13 $m_e$ is determined from the slope of $\Delta E(B)$. The upper inset is the semi-logarithmic Dingle plot from which the Dingle temperature $T_D$ = 8.5 K is obtained.}
\end{center}
\end{figure}

The Dingle temperature can be determined from the semi-logarithmic plot shown in the upper inset of Fig. 10 for three different temperatures. The dashed lines are a linear fit to the data and $T_D$ = 6.6 K is calculated from the slopes. With the parameters $m_c$ and $T_D$ fixed, the oscillation amplitude in the high-field limit is estimated from the data of Fig. 9 as $\Delta R_0$ = 5.04 $m\Omega$. The three parameters completely define the bulk SdH oscillation amplitude, which dominates the quantum oscillations above 15 Tesla, as function of magnetic field and temperature.

\begin{table}[h]
\caption{Comparison of the relevant parameters of bulk and surface SdH oscillations of Bi$_2$Se$_{2.1}$Te$_{0.9}$.}
\vspace{0.2cm}
\centering
\begin{tabular}{@{\hspace{0.25cm}} c @{\hspace{0.25cm}}|@{\hspace{0.25cm}} c @{\hspace{0.25cm}}|@{\hspace{0.25cm}} c @{\hspace{0.25cm}}|@{\hspace{0.25cm}} c @{\hspace{0.25cm}}}
\hline
&$\Delta R_0$ ($m\Omega$)&$m_c/m_e$&$T_D$ (K)\\
\hline
Bulk&5.04&0.34&6.6\\
Surface&2.6&0.13&8.5\\
\hline
\end{tabular}
\end{table}

In the low-field range, the SdH oscillations are determined by the topological surface states. A similar evaluation within the LK theory, restricted to below 10 Tesla, reveals the set of parameters for the SdH oscillations arising from the surface conduction. Some results of the LK analysis, based on data displayed in Fig. 9b, are shown in Fig. 11. For the current sample, the parameters determined for surface conduction are $\Delta R_0$ = 2.6 $m\Omega$, $m_c$ = 0.13 $m_e$, and $T_D$ = 8.5 K. The parameters for bulk and surface quantum oscillations are compared and summarized in Table 1. Note that the oscillation amplitude $\Delta R_0$ of the bulk SdH oscillations is larger by a factor of 2 as compared to $\Delta R_0$ of the surface sates, explaining the domination of bulk oscillations at higher magnetic fields. However, the cyclotron mass $m_c$ of the bulk oscillations is also significantly larger than that of the surface conduction resulting in a faster exponential decay at lower fields and higher temperatures. Although the Dingle temperature $T_D$ is slightly lower in the bulk, the product $m_c\cdot T_D$, which determines the exponent of $\lambda_D$ in equ. (1), is still larger and the bulk oscillations decrease more rapidly upon decreasing magnetic field. Therefore, the SdH oscillations are dominated by surface states in the low-field range. As an example, we show in Fig. 12 the oscillation amplitudes for both, surface (frequency $F_1$) and bulk (frequency $F_2$) transport, at 5 K calculated with the parameters from Table 1. It is obvious that, with increasing magnetic field, there is a crossover from surface dominated to bulk SdH oscillations. For example, at 5.85 Tesla (data shown in Fig. 10) the ratio of surface and bulk oscillation amplitudes is about 9, demonstrating the dominance of quantum oscillations from topological surface states at this field. At 14 Tesla, both oscillation amplitudes are equal resulting in the strongest interference. Below and above this crossover field, surface and bulk oscillations can well be separated, as shown in the frequency analysis above (Figs. 4 and 5).

\begin{figure}
\begin{center}
\includegraphics[angle=0,width=3in]{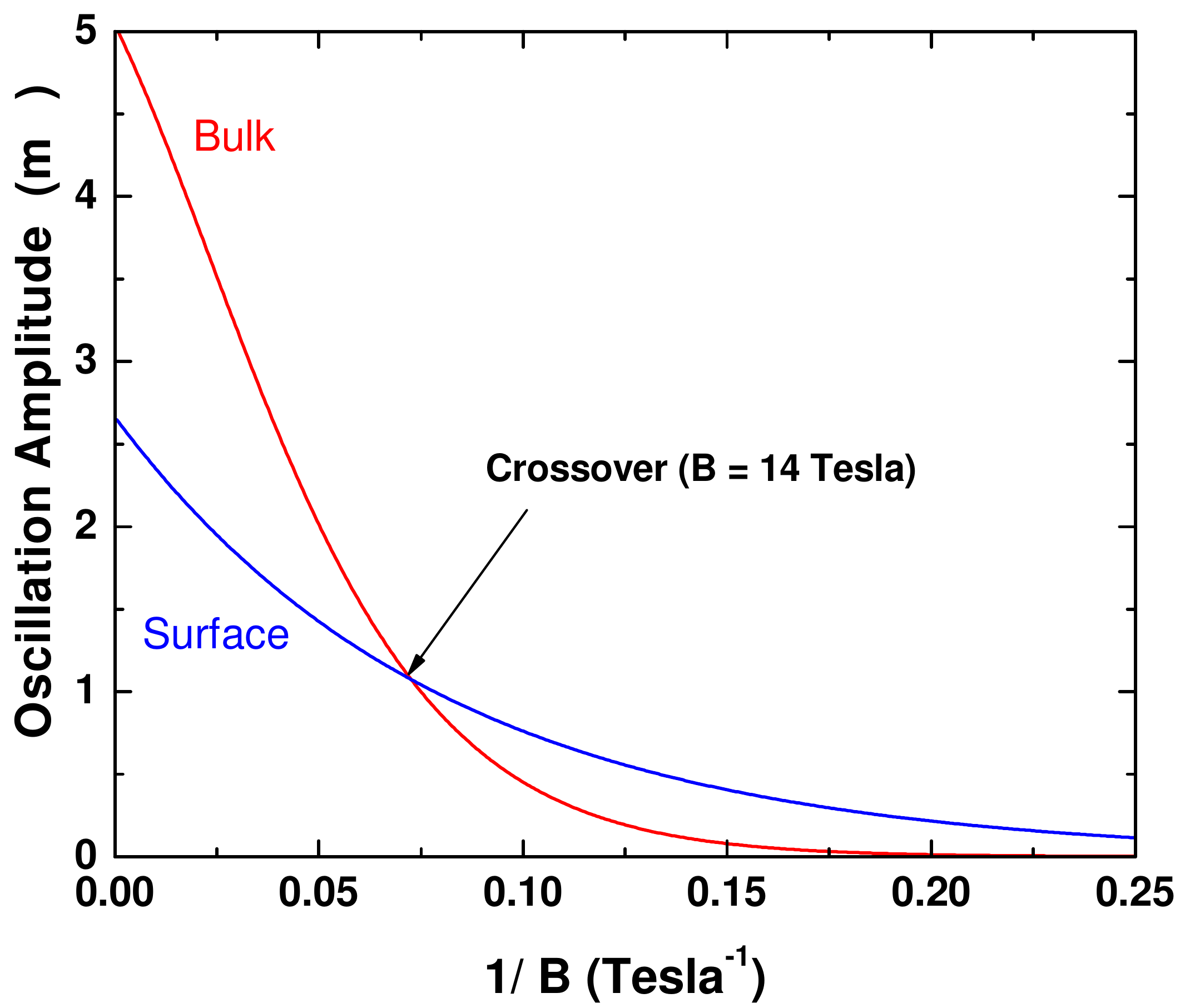}
\caption{Amplitudes of bulk and surface SdH oscillations as function of inverse magnetic field at 5 K. The crossover point of 14 Tesla is indicated in the figure. The curves have been calculated from the LK theory (equ. 1) with the parameters listed in Table 1.}
\end{center}
\end{figure}

\section{Summary and Conclusions}
The question of how to detect topological surface states utilizing different experimental techniques has been discussed in recent years. The study of the electronic excitation spectrum in ARPES measurements has revealed the Dirac-like dispersion filling the semiconducting bulk gap in various systems with strong spin-orbit interactions. Alternatively, magnetotransport investigations have shown the characteristics of two-dimensional conduction arising from topological surface states in the angle dependence of the frequency of Shubnikov-de Haas oscillations. This method was believed to work well as long as the bulk states do not contribute to the conduction \cite{taskin:11, qu:11, analytis:11, eto:11, cao:13}. In metallic topological compounds, however, the interference of SdH oscillations from surface and bulk states makes it very difficult to separate and study surface and bulk oscillations.

In the current example, metallic Bi$_2$Se$_{2.1}$Te$_{0.9}$ with hole-type carriers, Shubnikov-de Haas oscillations have been observed in magnetic fields up to 35 Tesla. Two characteristic oscillation frequencies, $F_1$ and $F_2$, can be clearly distinguished and attributed to oscillations from surface and bulk states, respectively. The character of the surface and bulk carriers is determined from the angle dependence of the SdH oscillations and the derived Berry phases. It is demonstrated that both oscillations can be separated whereas the topological surface states dominate in the low-field range and the bulk oscillations increase in relative weight at higher magnetic fields. The main origin of this separation is found in the different cyclotron masses ($m_c^{bulk}/m_c^{surf}\approx3$) which causes the bulk oscillations to decay (exponentially) more rapidly if the magnetic field is decreased. At a temperature of 5 K, the crossover from bulk to surface dominated quantum oscillations upon decreasing field is found at a critical value of $B_c$=14 Tesla.

The results of this study pave the way to study topological materials with bulk metallic properties using magnetoconductance measurements. They show that SdH oscillations from topological surface states can be detected even when the Fermi energy cuts through the valence band and the bulk transport properties are metallic. The conditions for a successful separation of surface and bulk SdH oscillations have been identified. The key parameter is the difference of the cyclotron masses $m_c$ which have a profound effect on the oscillation amplitudes as a function of magnetic field. According to the Lifshitz-Kosevich theory, the oscillation amplitude decreases exponentially with the inverse magnetic field and the exponent is determined by $m_c$. In the current example, Bi$_2$Se$_{2.1}$Te$_{0.9}$, the field ranges where bulk and surface oscillations dominate, are well separated and the analysis of the quantum oscillations can be conducted at high and low fields revealing the fundamental parameters of bulk and surface oscillations, respectively. Other topological systems with bulk metallic conduction are expected to show similar properties and may be analyzed following the procedure outlined in this work.

\section*{Acknowledgement(s)}
This work is supported in part by the T.L.L. Temple Foundation, the J.J. and R. Moores Endowment, the State of Texas through TCSUH, the US Air Force Office of Scientific Research, and at LBNL through the US Department of Energy. V. M. acknowledges support from the Bulgarian Science Fund, project FNI-T-02/26. A portion of this work was performed at the National High Magnetic Field Laboratory, which is supported by National Science Foundation Cooperative Agreement No. DMR-1157490 and the State of Florida. The work at Idaho National Laboratory is supported by Department of Energy, Office of Basic Energy
Sciences, Materials Sciences, and Engineering Division and
through Grant No. DOE FG02-01ER45872

\end{document}